\documentclass[prb,superscriptaddress,showpacs,twocolumn]{revtex4}

\usepackage{amsmath,amssymb,graphicx,color,bm}
\usepackage{ulem}

\begin{document}

\title{Nonlinear Bloch-waves and current states of exciton-polariton condensates}

\author{I.Yu. Chestnov}
\affiliation{Vladimir State University named after A. G. and N. G. Stoletovs, Gorkii St. 87, Vladimir, Russia}

\author{A.V. Yulin}
\affiliation{ITMO University 197101, Kronverksky pr. 49, St. Petersburg, Russian Federation}

\author{A.P. Alodjants}
\affiliation{Vladimir State University named after A. G. and N. G. Stoletovs, Gorkii St. 87, Vladimir, Russia}
\affiliation{ITMO University 197101, Kronverksky pr. 49, St. Petersburg, Russian Federation}

\author{O.A. Egorov}
\affiliation{Institute of Condensed Matter Theory and Solid State Optics, Abbe Center of Photonics, Friedrich-Schiller-Universit\"at Jena, Max-Wien-Platz 1, 07743 Jena, Germany}

\date{\today}

\begin{abstract}

The formation of nonlinear Bloch states in open driven-dissipative system of exciton-polaritons loaded into a weak-contrast 1D periodic lattice is studied numerically and analytically. The condensate is described within the framework of mean-field theory by the coupled equations for the order parameter and for the density of incoherent excitons. The stationary nonlinear solutions having the structure of Bloch waves are studied in detail. It is shown that there is a bifurcation leading to the appearance of a family of essentially nonlinear states. The special feature of these solutions is that its current does not vanish when the quasi-momentum of the state approaches the values equal to the half of the lattice constant. To explain the bifurcations found in numerical simulations a simple perturbative approach is developed. The stability of the nonlinear states is examined by  linear spectral analysis and by direct numerical simulations. An experimental scheme allowing the observation of the discussed nonlinear current states is suggested and studied by numerical simulations.
\end{abstract}

\maketitle

\section{Introduction}

Artificial periodic materials have been attracting attention of scientists working in different areas of physics including fundamental problems of photonics, material science, condensed matter and solid state physics \cite{PhotCryst, Nat.Nano.8.625.2013, PolInPerStr}. One of the reasons of the interest is that elementary excitations in these media have rich dispersion properties and, therefore, provide a convenient tool for control over the wave propagation. More generally, modification of a band structure of a given physical system is a vital practical problem of the design of new micro- and nanostructures with unusual features \cite{Nat.Comms.6.8682.2015}. Besides the effective control over the  dispersion of electromagnetic waves, the nonlinear properties of the waves in artificial media based on metamaterials and plasmonics are also in the focus of modern research \cite{MLapine14, MKauranen2015}.

One of the most important and promising trends in nonlinear photonics is the investigation of coherent hybrid light-matter excitations possessing unique optical properties \cite{PolInPerStr}. For instance, hybrid states of excitons and photons, known as exciton-polaritons, occur inside a high Q-factor  semiconductor microcavities operating in the strong coupling regime \cite{Weisbuch92}.  Extremely small effective mass of these composite bosons enables the observation of a high temperature nonequilibrium condensate of exciton-polaritons \cite{Kasprzak06, Deng2010}. Nowadays semiconductor microcavity supporting exciton polaritons is a very promising platform for the investigation of the spontaneous build up of macroscopic coherence in artificial periodic lattices \cite{Lai2007, Cerda10, Tanese2013, Winkler16}.

Important features of  such physical systems  are connected with their strong nonlinearities arising from two-body exciton-exciton interactions which crucially affect the properties of the condensate.
For instance, it has been revealed that in the presence of periodic modulation  self-localized states  known as gap solitons  can form in polariton systems \cite{Tanese2013} and in atomic condensates  \cite{Eiermann04} with either attracting or repulsive nonlinear interactions between the quasi-particles. The problem of the formation of the coherent states with strong nonlinear interactions looks even more intriguing in the light of possible superfluidity in the atomic \cite{Leg} and the polariton \cite{Carusotto2004, Amo2009} systems. Thus the study of the dynamical and spectral properties of the states forming in nonlinear  periodic systems is a task having fundamental importance.

In the case of relatively weak nonlinearity (for small number of particles in the condensate) the problem can be approached perturbatively when the solution is sought in the form of the expansion over the basis of the Bloch modes calculated for the corresponding linear problem and treating the nonlinearity as a perturbation, see e.g. \cite{Ziman}. For instance, in \cite{Charukhchyan2014} a powerful \textbf{kp}-method originally proposed in solid state physics (see e.g. \cite{Callaway}) is used to describe atomic \textit{Bose-Einstein condensate} (BEC) in optical lattices.
Naturally,  within the approaches of such a kind all the modes with quasi-momentum lying at the edge of \textit{Brillouin zone} (BZ) are currentless \cite{Ash}. However this conclusion cannot be generalized for the case of strong nonlinearity when the discussed perturbation approach fails.

The nonlinear band structure obtained beyond perturbative methods  was discussed  for conservative atomic BEC placed in 1D optical lattice \cite{VKonotop2004, Kramer2003, Diakonov2002, Mueller2002, Wu2003, Pethick2003, Seaman2005, Danshita2007}. In particular, it was shown that two-body (atom-atom) interaction leads to the formation of the loops in the dispersion characteristics of the nonlinear states (so-called ``swallow tails'') both at the center and at the edges of BZ. {Dispersion characteristics of nonlinear waves in this case are specified as  the dependencies of frequency of the these states versus quasi-momentum for the fixed average density of the condensate.}

The most intriguing consequence of the bifurcation leading to the formation of the loop structure is that this bifurcation gives birth to the nonlinear Bloch states that possess non-zero current even when their quasi-momentum is at BZ boundary \cite{Diakonov2002}. Note that the existence of such current is impossible in the linear case according to the Bloch theorem. Nevertheless, there is no contradiction here, since the Bloch theorem was formulated and proved for the problem of  one-particle motion in periodic potentials (see e.g.\cite{Ash}), whereas BEC is strongly nonlinear system. Note, that the existence of the flow at the boundary of BZ was interpreted as consequence of superfluid behavior of a condensate in a lattice \cite{Wu2003, Mueller2002}. The evidence of similar supercurrent states has been recently observed in atomic BEC with a ring geometry \cite{Wright2013,Eckel2014}.

The linear stability properties of the explored nonlinear states of atomic BEC, including the non-zero current states at the the BZ edges, have been intensively studied \cite{Wu2003, Pethick2003}. The regions of both the energetic (Landau) and dynamical instabilities have been identified for such essentially conservative systems. Note, as it was pointed in \cite{Wu2003}, the former stability criteria coincides with the Landau conditions for the breakdown of superfluidity. It was also mentioned that the periodic lattice enhances the dynamical instability resulting in the exponential growth of the perturbations.

The novelty of our paper is that it addresses open-dissipative (i.e. non-equilibrium) exciton-polariton condensates placed in periodic lattices.  {Contrary to the conservative bosonic systems \cite{Pethick_BEC} the exciton-polaritons are subject to rapid radiative decay, and their population must be supported by an external pump. We consider the case of the non-resonant pumping, which implies the existence of a reservoir of hot excitons that can relax in energy populating the polariton condensate. This nonequilibrium driven-dissipative behaviour is an important intrinsic feature of the exciton-polaritons in semiconductor microcavities and this crucially affects the nonlinear dynamics of both homogeneous \cite{Liew2015} and periodically modulated polariton systems \cite{X.Ma_PRB}.}

We perform comprehensive theoretical studies of the nonlinear Bloch states including the analysis of their dynamical stability and numerical simulations showing the effect of finiteness of the realistic system on the discussed solutions. For the sake of clarity and simplicity we restrict our consideration to a 1D case. {As it will be shown, an open-dissipative nature of polariton system gives rise to a number of interesting effects that are not possible to observe in their conservative counterparts.}

The paper is organized as follows.
To describe polariton condensate properties (see Sec.~\ref{Ch:Model}) we use an open driven-dissipative model that accounts for the particle exchange between the condensate and the excitonic reservoir formed by a non-resonant pump \cite{Wouters07}.
Sec.~\ref{Ch:BandStructures} deals with direct numerical simulations and the analysis of the eigenfrequency-quasimomentum diagram of the nonlinear states. Bifurcation and the stability of the nonlinear current states are discussed as well.  Then in Sec.~\ref{Ch:AnalytTheory} we develop a simple perturbation theory explaining the bifurcations of the nonlinear current states from a Bloch state at the boundary of the BZ.
In Sec.~\ref{Ch:DynamicsCurrent} we   investigate the formation and decay of the predicted  current states in the finite systems with inhomogeneous pumping of a special shape.
The protocol of experimental observation of the nonlinear Bloch current states at BZ boundary is proposed in this section.

\section{Theoretical model} \label{Ch:Model}

To describe  exciton-polariton  condensates in the mean field approximation  assuming the spontaneous formation of the condensate, we adopt a well known open-dissipative Gross-Pitaevskii (GP) model, that describes incoherently pumped condensate coupled to an excitonic reservoir. We describe polariton order parameter $\Psi(x,t)$ and the exciton reservoir density $n(x,t)$ by coupled GP-type and rate equations respectively, cf.~\cite{Wouters07, X.Ma_PRB},
\begin{subequations}
\label{main}
\begin{eqnarray}
i \hbar \frac{\partial \Psi}{\partial t} = \left[ -\frac{\hbar ^2}{2 m} \frac{\partial ^2}{\partial x^2} + V(x) + g_c \left| \Psi \right|^2 + \right.  \nonumber \\
\left. \frac{i \hbar}{2} (Rn-\gamma_c) + g_{R} n \right] \Psi, \\
\frac{\partial n}{\partial t} = P(x) - \gamma_R n - R n  \left|\Psi \right|^2,
\end{eqnarray}
\end{subequations}
where $\gamma_c$ and $\gamma_R$ are the condensate and the reservoir dumping rates, $g_c$ and $g_R$ are coefficients of nonlinearity due to polariton-polariton and polariton-reservoir interaction. The reservoir of incoherent polaritons is formed by an external pump with a spatially dependent particle generation rate $P(x)$. The parameter of the coupling between the reservoir and the condensate is denoted as $R$. The value of  $V(x)= V_0 \cos(\beta x)$ is an external lattice potential, where $V_0$ characterizes the depth of the lattice, $\beta=2\pi/l$ is the lattice wave vector, and $l$ is the lattice period.
We expect that nonlinearity affects polariton band structure if the term $g_c |\Psi|^2$ is comparable or higher than the potential depth $V_0$. So we consider shallow  polariton lattices which could be formed for instance by surface acoustic waves \cite{Cerda10, Lima2006}.

Throughout the  paper we use following parameters, which agree with experimental data \cite{Roumpos2011}: $\gamma_c =0.33$~ps$^{-1}$, $\gamma_R = 0.495$~ps$^{-1}$, $g_c = 6 \times 10^{-3}$~meV$\mu$m$^2$, $g_R=2g_c$, $R=0.01$~ps$^{-1}$$\mu$m$^{2}$, $m=0.568$ meVps$^2$$\mu$m$^{-2}$, $l=8$~$\mu$m.

Note that in the model \eqref{main} we use an approximation of a  parabolic bare polariton dispersion. This is valid for the case of small lattice wave vectors $\beta$ (i.e. large periods $l$) when the edge of the Brillouin zone, $\beta/2$, locates at the bottom of the real polariton dispersion where it can be confidently approximated as parabolic. For our parameters $\beta \simeq 0.785$~$\mu$m$^{-1}$ that is in order of magnitude smaller than a typical value of Rabi frequency which determines the depth of the parabolic region of the bare polariton dispersion. We would like to notice that in a general case the high order  dispersion of bare polaritons as well as the existence of the upper polariton branch can significantly affect the dynamics of the polariton condensates.

In the absence of the external potential ($V_0=0$) and for homogeneous pumping $P(x)=P_0$ a steady-state homogeneous solution (HS) of the system~\eqref{main} coincides with that known for the planar cavity. For the sake of generality we also allow HSs with nonzero
quasi-momenta $k\neq0$,  which have the form of traveling waves
\begin{equation}
{\Psi _{   HS}}(x,t) = {\Psi_0}{e^{ - i\,{\mu_0}t + i{k}x}},
 \label{eq:HomSol}
\end{equation}
where the condensate energy is given by $\hbar \mu_0 \equiv \hbar \mu\left( {{k},{{\left| {{\Psi _0}} \right|}}} \right) = \left( {{{{\hbar ^2}} \mathord{\left/
  \right. \kern-\nulldelimiterspace} {2m}}} \right)k^2 + {g_c}{\left| {{\Psi _0}} \right|^2} + {g_R}{n_{0}}$.
The HS starts to form when the pump exceeds a threshold ${P_{th}} = {\gamma_c \gamma _R}{ \mathord{\left/  \right. \kern-\nulldelimiterspace} {{R} }}$ and becomes sufficient to compensate for all the losses.
 The coherent exciton-polariton and incoherent reservoir densities are given by ${\left| {{\Psi _0}} \right|^2} = {{{\kern 1pt} (P_0 - {P_{th}})} \mathord{\left/
 {} \right. \kern-\nulldelimiterspace} \gamma_c }$ and ${n_{0}} = {\gamma_c  \mathord{\left/ {} \right. \kern-\nulldelimiterspace} { {R{\kern 1pt} } }}$, respectively~\cite{Wouters07}.

\section{Steady-states and band-structure of the condensate}  \label{Ch:BandStructures}

This section comprises the results of the direct numerical simulations of steady-states forming in an open driven-dissipative polaritonic system with embedded 1D periodic lattice.
We restrict our consideration  to the nonlinear solutions having structure of Bloch waves $\psi(x,k)\exp(i k x )$ where $\psi(x,k)$ is a function with the period equal to the period of the external potential.
These  ``nonlinear Bloch modes'' (cf. with \cite{Wu2003}) are parameterized by the quasi-momentum $k$ and the average particle density.
First, in Subsection~\ref{Ch:BandStructures:Method}, we present the numerical method used for the simulations and discuss the nonlinear frequency-momentum band structure of the nonlinear exciton-polariton condensate. Modification of the band structure and formation of the loops for different pumping rates are studied in \ref{Ch:BandStructures:Loops}.
Then, in Subsection~\ref{Ch:BandStructures:Biffurcat}, we focus on the nonlinear current states bifurcating from the Bloch mode at the boundaries of Brillouin zone.

\subsection{Frequency-momentum diagram} \label{Ch:BandStructures:Method}

First we consider steady-state solutions of the system~\eqref{main} in the form of the nonlinear Bloch waves and build respective frequency-momentum diagram for an idealized case of a spatially homogeneous pumping $P(x)=P_0$  and infinite lattice.
For this aim we are looking for solutions in a general form
\begin{equation}
\label{stat_sol}
\Psi(x,t) = \psi(x,k) e^{ikx - i \mu t},   \ \ \ \ n(x,t) = f(x,k),
\end{equation}
where $k$ and $\mu(k)$ are a quasi-momentum and an eigenfrequency   of a Bloch wave given by the condensate $\psi(x,k)$ and the reservoir $f(x,k)$ components.
Since the nonlinear Bloch waves assumed to be periodic with the period $l$ they can be expanded into a basis of spatial harmonics of the external periodic potential:
\begin{equation}
\label{expancion}
\psi(x,k) = \sum^{N} _{j=-N} \psi_{j} e^{i\cdot j \beta x}, \ \ \ \ f(x,k) = \sum^{N} _{j=-N} n_{j} e^{i\cdot j \beta x},
\end{equation}
where $N$ is a number of spatial harmonics which is, of course, finite in numerical simulations.
Substituting \eqref{stat_sol} and \eqref{expancion} into \eqref{main} and collecting terms with equal exponential factors one obtains $4N+2$ equations with $4N+3$ unknowns (including the frequency $\mu$). A missing equation can be found from the stationarity condition for the entire condensate density ($\partial \left| \Psi \right|^2/ \partial t  = 0$) which is equivalent to the condition $ \rm{Im} \left(\mu\right) =0$. Similar method has been successfully applied for the calculations of the energy band structure of the conservative atomic BEC \cite{Pethick2003}.

\begin{figure}
\includegraphics[width=0.48\textwidth]{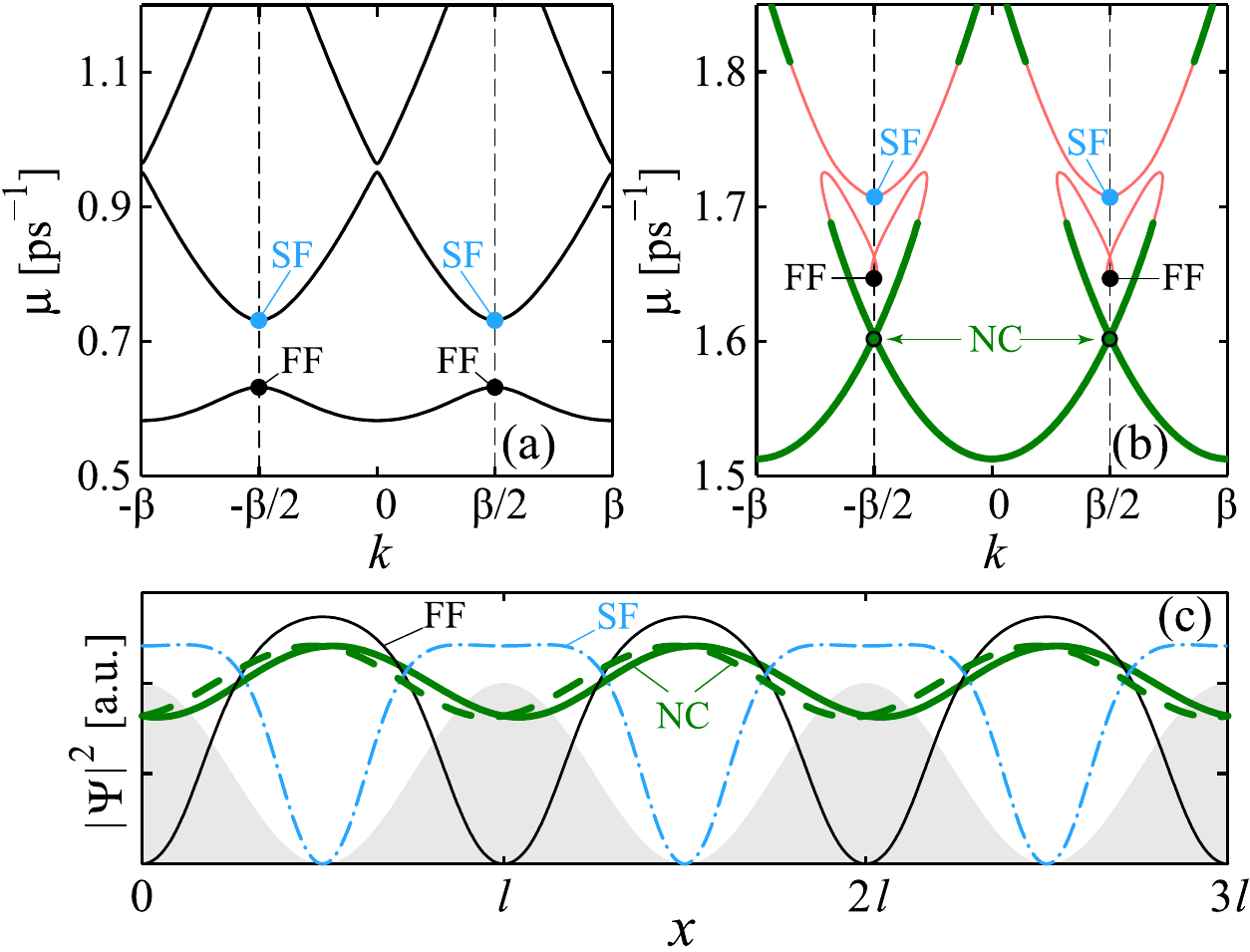}
\caption{(Color online) Frequency-momentum diagram of the exciton-polariton condensate in a one-dimensional periodic lattice ($V_0/\hbar=0.1$~ps$^{-1}$) plotted for different pumping rates. (a) Linear  band structure in the vicinity of the condensation threshold $P_{th}=16.335$~$\mu$m$^{-2}$ps$^{-1}$. (b) The formation of the loops in the nonlinear dispersion at the boundary of BZ for $P_0=50$~$\mu$m$^{-2}$ps$^{-1}$. Green lines correspond to dynamically stable solutions. (c) Spatial profiles ($|\Psi|^2$) of the nonlinear Bloch states at the BZ edge, namely  FF, SF and NC states, shown in the panel (b). The dashed and the bold solid green lines correspond to two NC-states describing two oppositely directed polariton flows. The shaded area mimics periodic potential $V(x)$. }
\label{fig_BandStr}
\end{figure}

This approach allows to determine profiles of the nonlinear Bloch waves and their frequencies $\mu$ for different quasimomenta $k$ and, thus, to reproduce the entire  frequency-momentum dispersion $\mu(k)$ of the exciton-polariton condensate. 
{For our further analysis it is necessary to extend the essentially linear conception of dispersion relation typical to conservative systems to the nonlinear driven-dissipative case. Under the nonlinear dispersion we understand eigenfrequency-momentum dependence $\mu(k)$ of the condensate steady-states sought in the form of the Bloch waves~(\ref{stat_sol}) for the fixed value of $P_0$. This is reasonable since in the open-dissipative polaritonic systems the density of the condensate is determined by the energy balance between the pump and losses. For further analysis we introduce the condensate density averaged over lattice cell, i.e. $I=\frac{1}{l}\int_{-l/2}^{l/2} \left|\Psi(x)\right|^2 dx$, which is equivalent to summation over discrete Fourier components $I=\sum^{N} _{j=-N} |\psi_{j}|^{2}$.}

In the vicinity of the threshold ($P_0\approx P_{th}$) the condensate density is small and nonlinear effects are negligible. In this case the dispersion $\mu(k)$ resembles a band structure of the dispersion of linear waves in a  periodic potential~\cite{Ash} -- see Fig.~\ref{fig_BandStr}(a). In the linear case the group velocity becomes equal to zero at the end of the BZ $k=\beta /2$ and thus these states carry no current, see \cite{Callaway}.
The growth of the pump increases the density $I$ of the condensate  evoking strong nonlinear modifications of the band structure, which admits bifurcation of additional ``extra'' nonlinear solutions of Eqs.~\eqref{main}.

Besides the nonlinear interaction of coherent polaritons [mediated by the term $g_c$ in Eq.~(\ref{main})] the incoherent reservoir also contributes to the nonlinear modification of the band structure. {This is one of the essential differences of the considered driven-dissipative system from its conservative counterpart\cite{Wu2003,Diakonov2002}.} Indeed, owing to the reservoir-polariton scattering [term $g_R$] the energy of the Bloch waves experiences an additional blue shift caused by the hot excitons. Therefore, a spatially-modulated reservoir induces an effective potential which has to be added to the external periodic lattice potential $V(x)$.  {Another important feature of driven-dissipative polariton  system is that the shape of the solution (and hence the shape of the total effective potential $V(x)+ g_R n(x) + g_c |\Psi(x)|^2$) is affected by the spatially distributed gain which is proportional to the reservoir distribution $n(x)$.}

All aforementioned mechanisms affect substantially both the shape and the stability of the Bloch waves resulting in the strong modification of the band structure. For instance, the lowest branch of the nonlinear dispersion characteristics can twist forming loops at the Brillouin zone edges -- see Fig.~\ref{fig_BandStr}(b). {Besides}  the shape of the second band in the vicinity of $k=0$ is also modified  {in the nonlinear regime}. However  {such modification occurs} only in the very narrow region close to the zone center (hundredth parts of BZ width) for chosen parameters of the system  {and thus it is not seen in Fig.~\ref{fig_BandStr}(b)}. Therefore here we focus on the investigation of the nonlinear Bloch waves in the vicinity of the BZ edges.

Typical profiles of the nonlinear Bloch states of the polariton condensate at the BZ edge are shown in Fig.~\ref{fig_BandStr}(c). Two of them emanate directly from linear Bloch states by means of the smooth variation of the pump rate $P_0$ parameter. Further we refer the solution families which originate from the top of the $1^{st}$ Bloch band as \textit{fundamental family }(FF), whereas \textit{second family} (SF) of the Bloch waves depicts the steady-states relating to the  $2^{nd}$ Bloch band.

Besides the ``conventional'' families (FF and SF) having their linear counterparts, the system possesses also essentially nonlinear Bloch waves directly associated with the formation of the dispersion loops. Remarkably that these Bloch states are characterized by the nonzero slope of the $\mu(k)$ dependence even at the boundaries of the Brillouin zone [see Figs.~\ref{fig_BandStr}(b) and (c)] and they have nonzero polariton current defined routinely \cite{Diakonov2002} as
\begin{equation}
\label{current}
j(x)= \frac{i\hbar}{2m}\left( \frac{\partial \Psi^{*}}{ \partial x} \Psi - \frac{\partial \Psi}{ \partial x} \Psi^{*} \right).
\end{equation}
Let us remark here that since $\mu(k)$ is the frequency-momentum diagram calculated for the fixed $P_0$ then $\partial \mu/\partial k$ is not a group velocity of the particles in open-dissipative system.

In what follows we refer the nonlinear Bloch states having a non-zero current at BZ boundaries as \textit{nonlinear current} (NC) states and consider them as a central object of our present study. {Further we focus on the shape of the dispersion loops forming close to the edges of BZ.}

{Note that the indication of the nonlinear modification of the energy band structure of the exciton polariton condensate in a periodic potential landscape was previously reported in the context of nonlinear Bloch oscillations of polariton current \cite{Flayac2011a,Flayac2011b}}.

\subsection{Dispersion loops at the edges of Brillouin Zone} ~\label{Ch:BandStructures:Loops}

\begin{figure}
\includegraphics[width=0.5\textwidth]{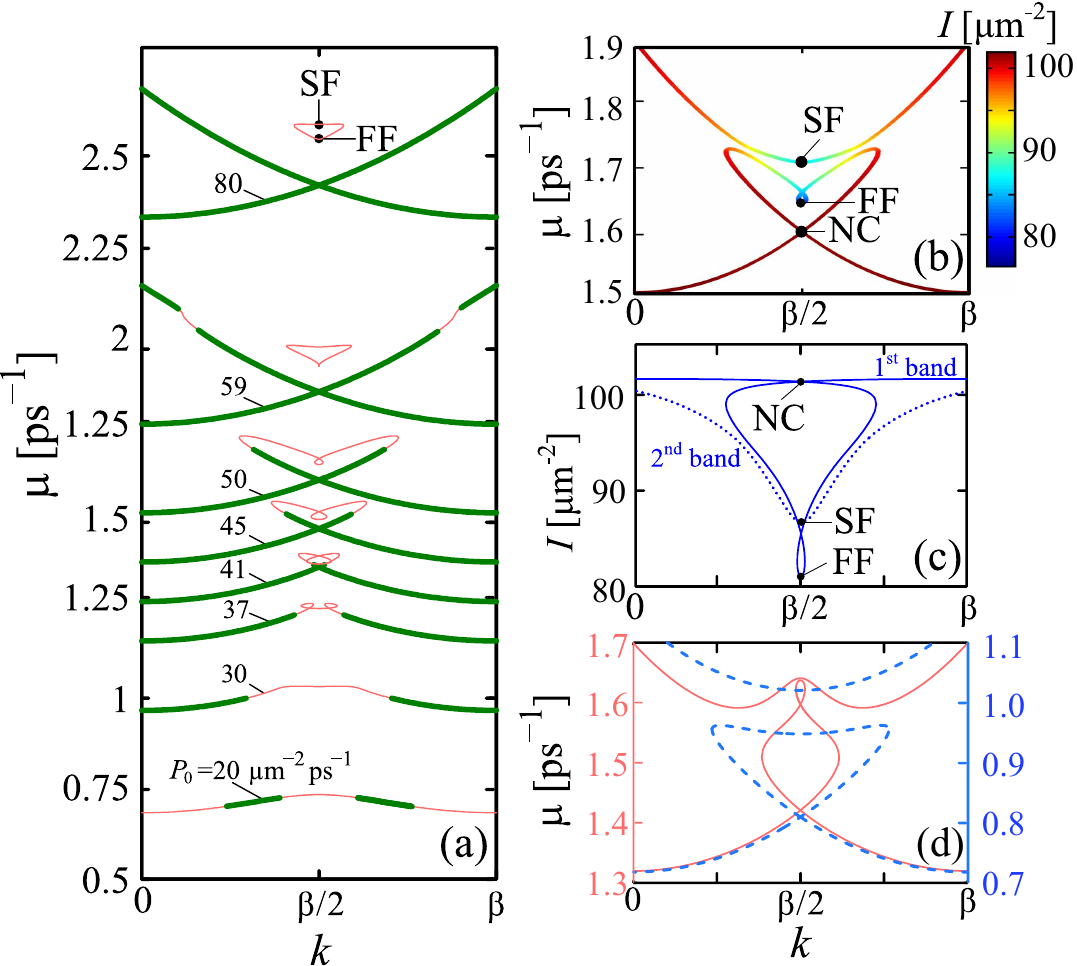}
\caption{(Color online)
(a) Bifurcation diagram showing the formation of the loops at BZ edge for fixed value of potential depth $V_0 / \hbar = 0.1$~ps$^{-1}$. Thick green sections correspond to dynamically stable states. (b) the dependency of the average density $I$ on the quasi momentum $k$ calculated for the fixed pump  $P_0=50$~$\mu$m$^{-2}$ps$^{-1}$ and  the same $V_0$ as in panel (a). The color of the curve shows how the average condensate density $I$ changes along the bifurcation diagram. (c) The same as in (b) but in the parameter plane of density $I$ and momentum $k$.  {(d) The eigenfrequency-momentum diagram of the solutions characterized by the same average condensate density $I=80$~$\mu$m$^{-2}$ for driven-dissipative polariton condensate (solid red, left vertical axis) and for conservative condensate (dashed blue, right vertical axis), $R=0$, $\gamma_c=0$ and $g_R=0$.} }
\label{fig_BandStr_BZ}
\end{figure}

Bifurcation diagram shown in Fig.~\ref{fig_BandStr_BZ}(a) illustrates the formation of the dispersion loops in the vicinity of the Brillouin zone boundaries for different values of the pumping rates $P_0$. 

In the nonlinear regime the curve on bifurcation diagram Fig.~\ref{fig_BandStr_BZ}(a) corresponding to the $1^{\rm{st}}$ Bloch band becomes flat at the vicinity of BZ boundaries [see curve $P_0=30$ $\mu$m$^{-2}$ps$^{-1}$]. Further, above a certain threshold, two loops emerge at the dispersion curve -- $P_0=37$~$\mu$m$^{-2}$ps$^{-1}$.  They move towards each other and tie into a knot at the zone boundary. Further increase of the pumping rate makes the edges of this knot stretch in the direction of the 2$^{nd}$ band forming a typical loop structure shown in Fig.~\ref{fig_BandStr}(b).

{In the limit of a very strong pumping $P_0$ (or vanishing potential contrast $V_0$) the first and the second bands merge [see curve $P_0=59$~$\mu$m$^{-2}$ps$^{-1}$ in Fig.~\ref{fig_BandStr_BZ}(a)], and the $\mu(k)$ dependence approaches parabolic shape [$P_0=80$~$\mu$m$^{-2}$ps$^{-1}$], which is typical for a ``free'' condensate (in the absence of periodic potential). Such a behaviour indicates that the influence of the external periodic potential on the dispersion properties becomes screened by the nonlinear effects. In Sec. IV we provide a description of the NC-state formation at the BZ edge, which, as will be seen, can be interpreted as a consequence of the nonlinear modification of the effective potential leading to the suppression of the 1$^{\rm{st}}$ order Bragg scattering of the polariton flow.}

Note, that 'NC' point in Figs.~\ref{fig_BandStr}(b) and \ref{fig_BandStr_BZ}(b) actually corresponds to two states with the oppositely directed currents,{ see also the dashed and the bold solid green curves in Fig.~\ref{fig_BandStr}(c).}
Besides these current states the conventional Bloch solutions (FF and SF) also survive  in the limit of very strong nonlinearity and form a closed curve in the dispersion [marked by FF and SF in Fig.~\ref{fig_BandStr_BZ}(a)].
These states resemble the strongly nonlinear Bloch waves found for the conservative case \cite{Pethick2003}, where these states were interpreted as arrays of equally spaced dark solitons, with one soliton per lattice period.

Note that the loops appearing in the dispersion resemble the so-called ``swallow tails'' predicted for the conservative atomic BEC \cite{Wu2003, Diakonov2002, Mueller2002, Pethick2003}. From the mathematical point of view swallow tail represents a special type of catastrophe described in the framework of catastrophe theory, see e.g. \cite{Kusmartsev}.  In many-body physical systems ``swallow tails'' occur as a solution  of  eigenvalue problem  for  the Hamiltonian containing nonlinear terms.  Strictly speaking  dispersion characteristics, i.e. dependence of the condensate eigenenergy versus momentum $k$,  in this case can be interpreted as bifurcation diagram characterizing physical system for a given value of condensate density.

It worth mentioning that  {in contrast to conservative case} the population of the nonlinear Bloch states {of non-resonantly pumped exciton-polariton condensate} is not a constant even along the same nonlinear dispersion curve calculated for a fixed value of pump $P_0$. In the driven-dissipative system the average density $I$ of the particular Bloch state is determined by the balance of losses and effective gain depending on the spatial overlap of the condensate distribution with the distribution of hot excitons. Since all states of the band have different spatial structures [see Fig.~\ref{fig_BandStr}(c)] they possess different densities and thus gain different blue shifts. An example of the distribution of the average polariton density over frequency-momentum structure is depicted in Figs.~\ref{fig_BandStr_BZ}(b) and (c).

{To compare discussed nonlinear dispersion characteristics of the exciton-polariton condensate with those obtained in the conservative limit (for  $R=0$, $\gamma=0$ and $g_R=0$), we plot the eigenfrequency-momenta diagram of the solutions \eqref{stat_sol} characterized with a fixed average condensate density $I$ -- solid (red) curve in Fig.~\ref{fig_BandStr_BZ}(d). Dashed (blue) curve corresponds to the conservative limit and represents the conventional ``swallow tail''. Note that in the limit of high particle density the polariton condensate eigenfrequency-momenta diagram plotted for the fixed $I$ also approaches parabolic shape which is accompanied with the merging of the two lowest bands.}

To check the dynamical stability of the solutions we use the standard Bogoliubov-de Gennes approach \cite{Li2015,Pethick2003,Ostrovskaya2014}, which implies perturbation of the solution \eqref{stat_sol} in the form
\begin{subequations}
\begin{eqnarray}
\psi(x,k) &=& e^{ikx- i\mu t} \sum_j \left[ {\psi_j} + \delta\psi_j e^{i\omega t + iq x}  \right.+ \\
&&  \left. \delta\varphi_j^{*} e^{-i\omega t - iq x}  \right]e^{i \cdot j \beta x} ,  \nonumber \\
n(x) &=& \sum_j \left[{n_j} + \delta n_j e^{i\omega t+ iqx} + \delta n_j^{*} e^{-i\omega t - iqx}  \right]e^{i \cdot j \beta x},
\end{eqnarray}
\end{subequations}
where $\omega$ is eigenfrequency of   small excitation with quasi momentum $q$. Substituting this solution into Eqs.~\eqref{main} and keeping only those terms which are linear in respect to small perturbations we obtain the linearized system of equations for all  {$\delta \psi_j$, $\delta\varphi_j$ and $\delta n_j$}. The Bloch state is unstable if the imaginary part of the eigenfrequencies  $\omega$ of the small excitations is negative for any values of $q$.

The domains of dynamical stability are shown by thick green curves in Figs.~\ref{fig_BandStr}(b) and \ref{fig_BandStr_BZ}(a).
The dynamical stability depends on the pump $P_0$ in a nontrivial way. In particular, the NC states at the BZ edges became stable provided that the pump amplitude overcomes some threshold value. For our parameter the threshold pump is approximately $P_0 \approx 40$ $\mu$m$^{-2}$ps$^{-1}$, see Fig.~\ref{fig_BandStr_BZ}(a).

Note that in the low-density limit (just above condensation threshold) the ground steady-states ($k\thickapprox0$) may be subject to modulational instability (see curve $P_0=20$ $\mu$m$^{-2}$ps$^{-1}$) as it happens in homogeneous polaritonic systems \cite{Ostrovskaya2014, Liew2015, Solnyshkov2014}. However this instability becomes suppressed in the strongly nonlinear regime.

\subsection{Bifurcation of nonlinear current states} ~\label{Ch:BandStructures:Biffurcat}

\begin{figure}
\includegraphics[width=0.48\textwidth]{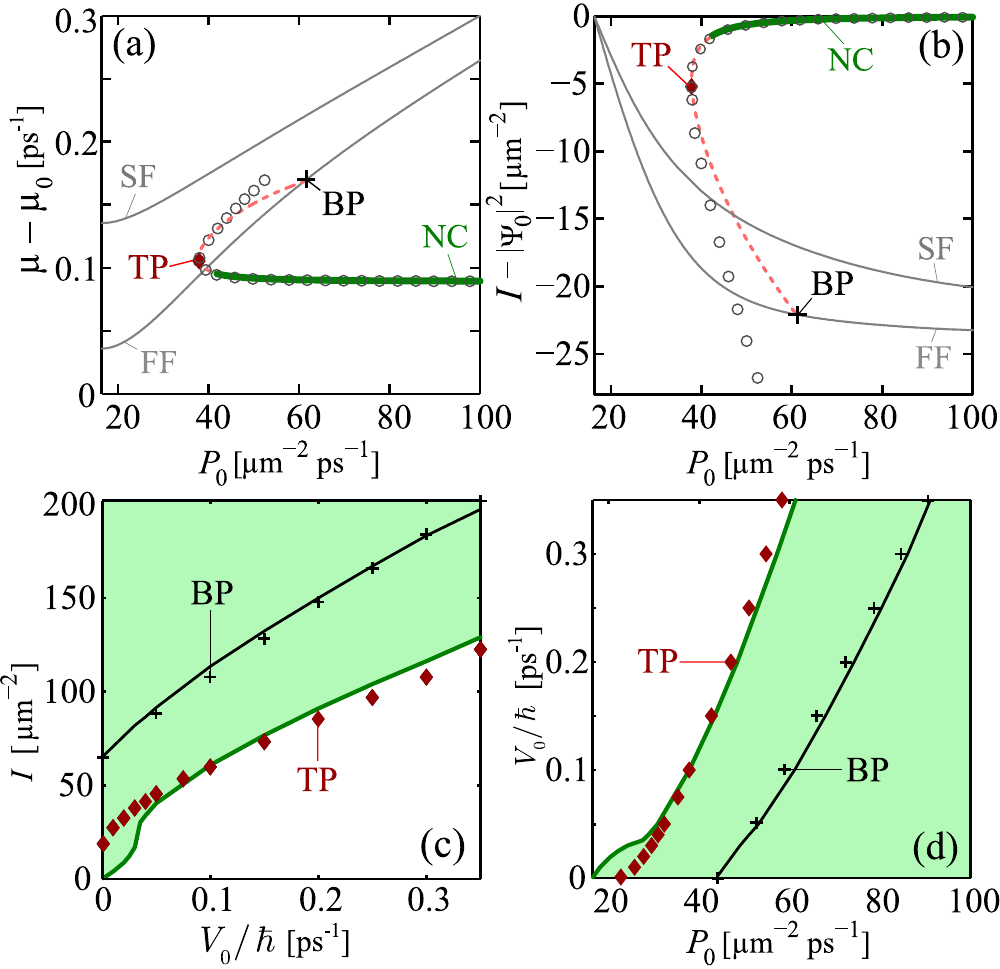}
\caption{(Color online) (a) Frequencies  $\mu$ of the nonlinear Bloch states at the BZ edges versus pumping rate $P_0$ for $V_0/\hbar=0.1$~ps$^{-1}$; (b) Averaged over a lattice cell density $I$  versus $P_0$ for the same potential depth.
Circles represent the analytical results~\eqref{Current_Freq_gr} and \eqref{DensityAnalytic_gr}.
(c) The existence domain (shaded area) of the current Bloch states in the parameter planes of the potential depth $V_0$ and the average density $I$. (d) The existence domain as in (c) but in the parameter plane of the pumping rate $P_0$ and the potential depth $V_0$. Thin solid lines show the bifurcation point (BP) and the turning point (TP) of the current state branch obtained by means of direct numerical simulation of the origin model \eqref{main}. Red diamonds corresponds to analytical results for the TP obtained from \eqref{DensityAnalytic_gr}, whereas the crosses show the BP obtained from the criteria Eq.~\eqref{BP_condition}.
}
\label{fig_bifurc}
\end{figure}

One of the remarkable consequences of strong modification of exciton-polariton condensate eigenfrequency-momentum diagram is the appearance  of the nonlinear Bloch states associated with a nonzero polariton flow $j\neq0$ at the edges of Brillouin zone.  {The formation of these states occurs only in the nonlinear regime and this implies that the 1$^{\rm{st}}$ order Bragg scattering of the polariton flow becomes effectively suppressed by the nonlinearity.} In this subsection we study these nonlinear Bloch states in detail.

Numerically calculated bifurcation diagrams Figs.~\ref{fig_bifurc}(a) and (b) elucidate behaviour of the nonlinear Bloch states at the BZ edge, $k=\beta/2$, as a function of pumping rate $P_0$ for a fixed potential amplitude $V_0/\hbar = 0.1$ ps$^{-1}$.
To show the influence of the lattice it is more instructive to examine how the parameters of the condensate placed in a periodical potential deviate from the parameters of the condensate in a spatially uniform system. We start with the frequencies of the condensates.
The dependency of the frequency difference $\mu-\mu_0$ on the pump is shown in panel (a) of Fig.~\ref{fig_bifurc}. Panel (b) of Fig.~\ref{fig_bifurc} shows the deviation of the average density of the polaritons in periodic lattice $I$ from the density of the polaritons in spatially uniform system $|\Psi_0|^2$ as a function of the pump.

Our numerical simulations show that the nonlinear current Bloch states  bifurcate from the fundamental family ``FF'' [Figs.~\ref{fig_bifurc}(a) and (b)].
The NC solution branch tends first to the smaller values of pump just after the bifurcation point (BP). However, since the current states don't exist in the linear limit, the turning point (TP) appears on the bifurcation characteristics.  Our linear stability analysis shows that the state is unstable just after the bifurcation point [dashed curve in Figs.~\ref{fig_bifurc}(a) and (b)], but the solution becomes stable after the TP of the NC solution branch.

With a further increase of the pumping rate $P_0$ the current states reach asymptotic solution in the form of ``free'' propagating particles given by the Eq.~\eqref{eq:HomSol}.  {It means that, when a nonlinearity gets stronger, the external potential becomes effectively screened. This interpretation agrees with the parabolic nonlinear dispersion under very strong pumping, as it is described in the previous subsection. }

We have performed an extensive numerical analysis of the system and found that the stable nonlinear current states exist in the wide range of the system's parameters. Figures~\ref{fig_bifurc}(c) and (d) show both the turning point (TP) and the bifurcation point (BP) of the solution versus the key system parameters, namely, the depth of the periodic potential $V_0$ and the pumping rate $P_0$. Shaded areas show the existence domains of the current states. We mention that outside this region [i.e. above 'TP' line in Fig.~\ref{fig_bifurc}(d)] the system possesses a solution in the form of persistent macroscopic oscillations, as it was shown recently by some of the authors~\cite{X.Ma_PRB}.

\section{Analytic description of nonlinear current states} ~\label{Ch:AnalytTheory}

In this  section we give an analytic description of  nonlinear Bloch states of exciton-polaritons loaded into a weak-contrast 1D periodic lattice. Usually, in such class of problems analytical methods are  suitable only for special kinds of a potential such as elliptic \cite{Bronski2001} or Kronig-Penney  \cite{Seaman2005, Danshita2007} potentials.
In the section we develop a simple theory explaining the bifurcations of the nonlinear current states from a nonlinear Bloch state at the boundaries of the BZ. 
{Starting from the solution known in conservative system we successively include the main peculiarities of the considered driven-dissipative system, namely, the spatial distributed gain which determines the average density and the shape of the state and the reservoir induced blue shift. So, first}, in subsection~\ref{Ch:AnalytTheory:Gr0}, we find the polariton density of the nonlinear current states in the limit of negligibly small blue shift induced by the reservoir. To do this we use an energy balance condition. Then (in subsection~\ref{Ch:AnalytTheory:Gr}) we generalize this approach taking into account the modification of the potential due to the spatially dependent blue shift.

\subsection{Neglecting the blue shift of the reservoir ($g_R=0$) } ~\label{Ch:AnalytTheory:Gr0}

For methodical reasons we start with a simplified problem disregarding the blue shift of the condensate caused by the incoherent excitons ($g_R=0$). The effect of the blue shift is considered  in the next subsection.
Another assumption we make is that  { the nonconservative terms $Rn-\gamma_c$ are small compared to the conservative terms in the sense that the balance between the gain and the losses defines the average density of the polaritons but the structure of the solution is determined mostly by the conservative terms and the impact of the dissipative terms on the structure of the solution is negligible.} 
Knowing a solution for $\psi$ we can find the solution for $n$ containing the pump intensity as a parameter. Then using energy balance condition we obtain an equation for the intensity of the pump needed to support the solution.

It is known that in the conservative case for a harmonic potential of amplitude $V_0$ there is an exact nonlinear solution consisting of two spatial harmonics \cite{Bronski2001, Wu2003}.
\begin{equation} \label{trial_psi}
\Psi(x,t) = \left( A e^{i \beta x/2} + B e^{ - i \beta x/2}   \right) e^{  - i \mu t}.
\end{equation}
Substituting the trial function~\eqref{trial_psi} into the conservative part of the model (\ref{main}a) the respective frequency and amplitude can be easily found
\begin{subequations}\label{CurrentState_sol}
\begin{eqnarray}
\mu &=& \frac{\hbar \beta^2}{8m} + \frac{g_c}{\hbar} I, \label{Current_Freq} \\
A &=& \pm \frac{1}{\sqrt{2}}\left( I \pm \sqrt{I^2 - \left( V_0 / g_c \right)^2} \right)^{1/2}, \\
B &=& \mp \frac{1}{\sqrt{2}}\left( I \mp \sqrt{I^2 - \left( V_0 / g_c \right)^2} \right)^{1/2},
\end{eqnarray}
\end{subequations}
where $I \equiv |A|^2 + |B|^2$ is the average condensate density.
Then the condensate density distribution of the nonlinear Bloch state is given by the following simple expression
\begin{equation}
\label{CurrentStateProfile}
|\Psi(x)|^{2} = I - \frac{V_0}{g_c} \cos\left( \beta x \right).
\end{equation}
{Note that the shape of the condensate implies that the nonlinear effective potential $g_c |\Psi(x)|^2$ \textit{completely compensates} the linear external potential $V(x)$. It means that the scattering of the waves into higher harmonics is absent and the solution \eqref{trial_psi} is exact.}  It is worth mentioning that the solution~(\ref{CurrentState_sol}) exists only if the following criteria is fulfilled:
\begin{equation}
\label{criteria}
 I \geq \frac {V_0} {g_c}.
\end{equation}
It implies that the respective Bloch mode doesn't  {exist in the linear limit (for small $I$)} and occurs when the nonlinear contributions are strong enough.  {It can be understood from the fact that in the linear case the potential cannot be compensated by the nonlinear terms and thus the current state at the end of the Brillouin zone is impossible.}

\begin{figure}
\includegraphics[width=0.48\textwidth]{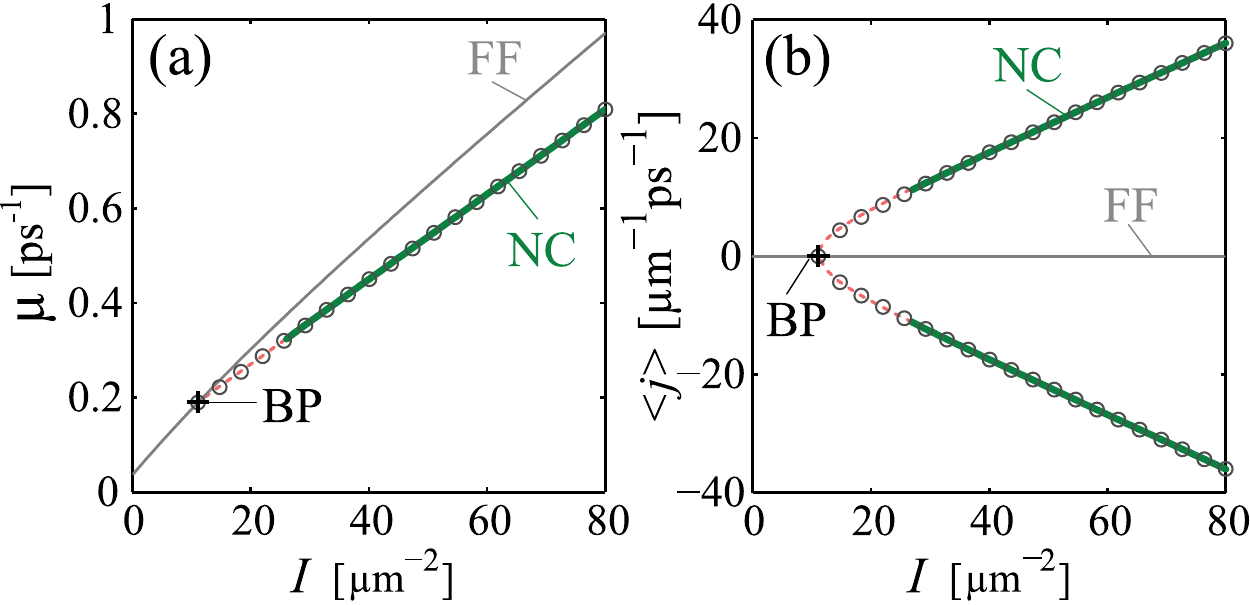}
\caption{(Color online) (a) Frequencies $\mu$  and (b) the averaged polariton current of the nonlinear Bloch states at the BZ edges versus their averaged density $I$. The nonlinear current (NC) Bloch states branch off from the fundamental family (FF) of the Bloch waves in the bifurcation point (BP). Two branches of the NC state correspond to oppositely directed polariton flows. Circles show the respective analytical results given by Eqs.~(\ref{CurrentState_sol}) and ~(\ref{current_BZ}). Thick solid (green) lines correspond to stable NC states. Parameters: $V_0/\hbar=0.1$~ps$^{-1}$ and $g_R=0$.}
\label{fig_bifurc_analyt}
\end{figure}

This nonlinear steady-state is characterized by a nonzero flow (\ref{current})  of the coherent polaritons averaged over a lattice period:
\begin{equation}
\langle j\rangle= \pm\frac{ \hbar \beta }{2m }  \sqrt{I^{2}- V_0^{2}/g_c^2  }. \label{current_BZ}
\end{equation}
To check the validity of this analytical approach we plotted both the eigenfrequency $\mu$ given by Eq.~(\ref{Current_Freq}) and the averaged polariton flow~(\ref{current_BZ}) versus the averaged densities $I$ of the Bloch states (see circles in Fig.~\ref{fig_bifurc_analyt}). These analytical results coincide with those obtained by means of direct numerical simulation (lines) of the model~(\ref{main}) for the case of $g_R=0$. These NC solutions branch off from the conventional Bloch state (FF) at the bifurcation point (BP) given by the condition~(\ref{criteria}).

The intensity of the solution with given quasi-momentum $k$ is a free parameter in the conservative case parameterizing the family of the solutions. In the dissipative case the situation is very different and the intensity of the solution is fixed by the balance of  gain and losses.
From Eqs.~\eqref{main} one can derive the balance equation in the form
\begin{equation}
\label{popul_cond}
\frac{\partial |\Psi(x)|^2}{\partial t} = \left( Rn(x) - \gamma_c \right) |\Psi(x)|^2 - \frac{\partial j(x)}{\partial x},
\end{equation}
where $j(x)$ is a flow of polaritons given by the Eq.~\eqref{current}. This energy balance condition allows to find the average density of the steady-state $I$ versus the pump $P_0$. To find the balance we integrate the equation (\ref{popul_cond}) over a period of the lattice $l$ and obtain a simple expression for the averaged density of any stationary state:
\begin{equation}
\label{al1}
I=\frac{1}{l \gamma_c} \int_{-l/2}^{l/2}Rn(x)|\Psi(x)|^2 dx,
\end{equation}
where the integral over a lattice period of the last term in Eq.~\eqref{popul_cond} is equal to zero for any Bloch states of the system.
The stationary reservoir density distribution can be found from Eq.~(\ref{main}b)
\begin{equation}
\label{al2}
n(x)=\frac{P}{\gamma_R+R|\Psi(x)|^2}.
\end{equation}
For our particular choice of the nonlinear Bloch states with $|\Psi(x)|^{2}$ given by~(\ref{CurrentStateProfile}) the integral~(\ref{al1}) can be calculated explicitly. This immediately gives the dependency of $P_0$ on $I$ for the stationary solution
\begin{equation} \label{DensityAnalytic}
P_0=\frac{\gamma_c I}{1-\frac{\gamma_R}{\sqrt{(\gamma_R + R I)^2 - (RV_0/g_c)^2}}}.
\end{equation}

To complete our analysis it is necessary to find the spatial distribution of the incoherent reservoir density. Due to the stimulated scattering and gain saturation effects, the nonlinear Bloch steady state~(\ref{CurrentState_sol}) imprints a periodic modulation of the reservoir density $n(x)$ given by Eq.~(\ref{al2}). In the limit of a relatively weak modulation of $n(x)$ it is reasonable to assume that the modulation of the reservoir density is harmonic and to truncate the expansion~(\ref{expancion}) at $N=1$:
\begin{equation}
\label{reserv_expan}
n(x)\approx n_0+ 2 n_{1} \cos(\beta x),
\end{equation}
To obtain this formula we also assumed that $n_{1}=n_{-1}$  which is justified by the shape of the external potential. For the polariton density distribution given by~(\ref{CurrentStateProfile}) it is possible to calculate the required Fourier components of $n(x)$  exactly
\begin{subequations} \label{n_estimation}
\begin{eqnarray}
n_0 &=& \frac{P_0}{\sqrt{(\gamma_R + R I)^2 - (RV_0/g_c)^2}}, \label{n_0} \\
n_1 &=& \frac{g_c}{RV_0}\left[ (\gamma_R + RI) n_0  - P_0 \right].\ \ \ \ \ \label{n_1}
\end{eqnarray}
\end{subequations}
Below we apply this analytical result for the description of the NC bifurcation branch of the Bloch states at the BZ boundary for realistic system parameters.

\subsection{Contribution of the reservoir-induced blue shift ($g_R\neq0$) } ~\label{Ch:AnalytTheory:Gr}

In real exciton-polariton systems the blue shift of coherent polaritons caused by incoherent excitons is strong and cannot be neglected.
To take this effect into account it is necessary to know the spatial distribution of the incoherent reservoir, which in simplest case can be approximated by~(\ref{reserv_expan}).
Thus, in the case of non-zero $g_R$, the Bloch wave induces an additional intensity-dependent periodic lattice $g_R n(x)$ mediated by the blue shift caused by the reservoir. The perturbation theory developed above can be further improved by taking into account the corrections induced by the first harmonic component $n_{1}$ of this additional potential. For this aim we consider an effective '$\cos$'-shaped lattice of amplitude $ V_0+2 g_R n_1$  which includes also a contribution of the reservoir modulation governed by~(\ref{n_1}).
All the computations presented above remain valid in this case with respective substitution $V_0 \rightarrow  V_0+2  g_R  n_1$.
Besides, the frequency of the current state solution acquires an additional reservoir-induced blue shift proportional to $n_0$
\begin{equation} \label{Current_Freq_gr}
\mu = \frac{\hbar \beta^2}{8m} + \frac{g_c}{\hbar} I + \frac{g_{R}}{\hbar} n_0.
\end{equation}
For the further analysis we also rewrite explicitly the expression~(\ref{DensityAnalytic}) connecting the pumping rate and the average density of the NC Bloch state for the corrected effective lattice
\begin{equation} \label{DensityAnalytic_gr}
P_0=\frac{\gamma_c I}{1-\frac{\gamma_R}{\sqrt{(\gamma_R + R I)^2 - \left( R(V_0+2 g_R n_1 )/ g_c \right)^2}}}.
\end{equation}

Solving Eqs.~(\ref{Current_Freq_gr}) and ~(\ref{DensityAnalytic_gr}) self-consistently and using \eqref{n_estimation} one can determine both the frequency $\mu$ and averaged density $I$ in the presence of reservoir-induced blue shift for arbitrary value of pumping rate $P_0$ -- see circles in Fig.~\ref{fig_bifurc}(a) and (b).
The analytical results accurately describe the NC states up to the turning point (TP), where the solution turns back towards the bifurcation point (BP).

The analytical dependence~(\ref{DensityAnalytic_gr}) also allows to estimate the existence domain of the NC solutions for different system parameters. The turning point can be considered as the lowest-pump boundary of the solution. The turning point can be found from the condition $\frac{\partial P}{ \partial I}=0$ resulting in a rather clumsy expression which, however, can easily be solved numerically. These semi-analytical results give a good approximation for the boundary of the NC states within the wide range of systems parameters, as it is shown by the red diamonds in Figs.~\ref{fig_bifurc}(c) and (d) (cf. with thick green curve obtained from numerical simulations).

We note that the analytical estimation [circles in Figs.~\ref{fig_bifurc}(a) and (b)] for the NC branch~(\ref{DensityAnalytic_gr}) becomes inaccurate in the vicinity of the bifurcation point (BP) where it branches off from the FF Bloch state. This discrepancies appear due to a strong deformation of the effective lattice potential induced by the incoherent reservoir $n(x)$ which can not be approximated any more by the simple harmonic dependency~(\ref{reserv_expan}).

Though the direct analytical description of the BP point seems to be difficult, we can apply  the criteria for the existence of the NC state, which requires a nonlinear screening of the effective lattice potential. Taking into account the blue shift caused by the reservoir we rewrite the criteria~(\ref{criteria}) at the bifurcation point in the form
\begin{equation}
   I_{BP} = \frac{ V_0 + g_R \Delta n }{ g_c } . \label{criteria_gr}
\end{equation}
Since the reservoir-induced periodic lattice is not necessarily '$\cos$'-shaped (\ref{reserv_expan}) we defined the modulation depth of the lattice as $ \Delta n \equiv (n_{\rm{max}}-n_{\rm{min}})/2$. To estimate this value we scrutinize the spatial profile of the FF Bloch state which has density maximum ($|\Psi_{\rm{max}}|^2$) at $x=l/2$ and zero density at $x=0$, -- see Fig.~\ref{fig_BandStr}(c).
Then the required estimation for $\Delta n$ can be obtained from the equation for reservoir population \eqref{al2}  as
 \begin{equation}
   \Delta n = \frac{ R P_0 |\Psi_{\rm{max}}|^2}{2\gamma_R(\gamma_R+R |\Psi_{\rm{max}}|^2)}. \label{delta_n}
\end{equation}
 We note that the maximal density in the FF state profile $|\Psi_{\rm{max}}|^2$ with good approximation can be related to its averaged over period density $|\Psi_{\rm{max}}|^2=2I$, at least, for our choice of system parameters.

In the vicinity of the BP, where the NC branch bifurcates from the FF states, we expect that the criteria~(\ref{criteria_gr}) is satisfied. Together with the estimation~(\ref{delta_n}) it immediately delivers the following condition for the pumping rate $P_{BP}$ and the averaged density $I_{BP}$ in the bifurcation point:
\begin{equation}
P_{BP}=  \frac{\gamma_R (g_c I_{BP}-V_0)(\gamma_R+2 R I_{BP})} {g_R R I_{BP}} . \label{BP_condition}
\end{equation}
Then, the parameters for the BP can be found from the condition~\eqref{BP_condition} with the use of the numerically calculated dependence of the density of the FF state $I$ on $P$.
Figs.~\ref{fig_bifurc} (d) and (c) show both the density and the critical value of the pumping rate at the BP for different modulation depths (black crosses). The analytical estimations for the bifurcation point (BP) are in good agreement with the results of the direct numerical simulations (black thin curve).

{Note that in the general case the shape of the reservoir induced potential is not strictly harmonic even far way form BP. It means that although  the condensate nonlinearity $g_c |\Psi(x)|^2$ aspires to compensate the effective potential $V(x) + g_R n(x)$, the total effective potential $V_{\rm{eff}}=g_c |\Psi(x)|^2+V(x) + g_R n(x)$ does not vanish completely. Actually only partial screening of the potential takes place. Our numerical simulations confirm it. However the current states at the edges of BZ still exist even in this case. }

\section{Excitation dynamics of the nonlinear current states} ~\label{Ch:DynamicsCurrent}

\begin{figure}
\includegraphics[width=0.9\linewidth]{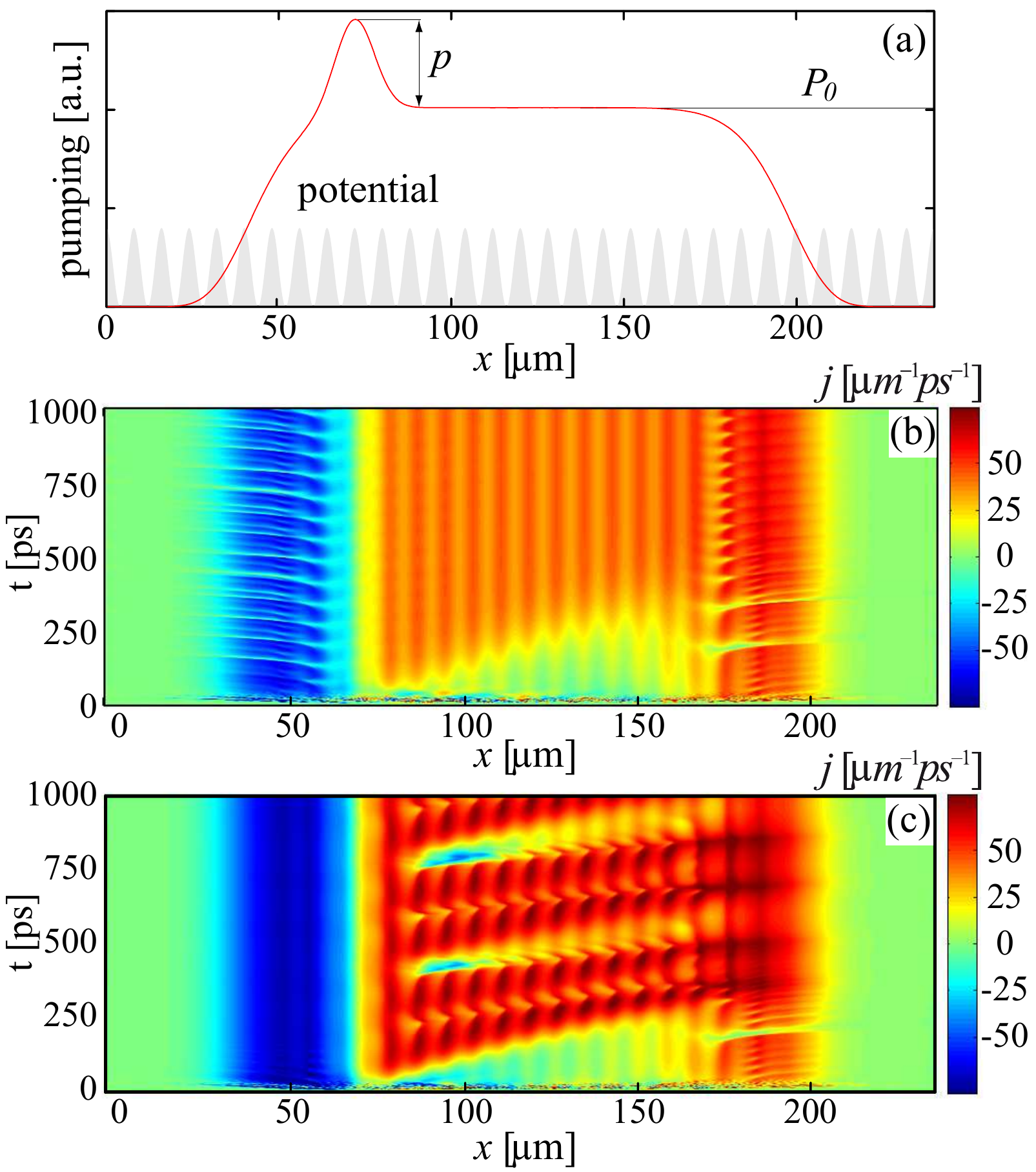}
\caption{(Color online) (a) The suggested scheme of the pump allowing for experimental observation of the current states. (b) Laminar dynamics of the coherent polaritons current $j$ for the pump peak amplitude $p=8$~$\mu$m$^{-2}$ps$^{-1}$ and background pumping rate $P_0 =50$~$\mu$m$^{-2}$ps$^{-1}$. (c) Turbulent dynamics of the polariton current for $p=16$~$\mu$m$^{-2}$ps$^{-1}$ and $P_0 =50$~$\mu$m$^{-2}$ps$^{-1}$.
}
\label{fig_curr_dynamics}
\end{figure}

The nonlinear Bloch states can exist and be stable in both conservative and dissipative systems. However it is not easy to suggest an effective method allowing to excite the states close to the BZ edge in the conservative systems. However one, perhaps exceptional, method was used for the observation of the hysteresis loops associated with energy loop structure in atomic BEC with a ring geometry \cite{Wright2013,Eckel2014}.  {In the case of the conservative systems the state must be created with good accuracy by the manipulation of the polaritons condensed into a ground state. Usually it is a difficult task.}

The advantage of the dissipative systems is that the nonlinear states are normally attractors and thus it is possible to design a system where such states can form from weak noise.  {Then to obtain the desired stationary state  it is sufficient to provide an initial distribution within the basing of attraction of the state.}
To examine the dynamical properties of the obtained nonlinear current states  we propose a numerical experiment  of the  condensation process starting from initial noise.

To obtain current state in a polariton condensate one needs a source of   current. For this goal we suggest to use a spatially inhomogeneous pump that could be exploited in real experiment with exciton-polariton condensate. One of the possible realization of such pumping is shown in Fig.~\ref{fig_curr_dynamics}(a). It consists of a long  plateau (tens of lattice periods)  with $P=P_0$ where the solutions similar to the solutions found for the infinite systems can form. The pump has a localized peak with an amplitude $p$ which plays the role of an injector providing a non zero current flowing into the plateau region. The injection of the polaritons creates the boundary condition needed for the formation of the current state. The sink for the polaritons at the other end of the plateau region is provided by usual losses.   The required distribution of the pump can be experimentally realized by using superposition of a gauss and a supergauss laser beams.
\begin{figure}
\includegraphics[width=1\linewidth]{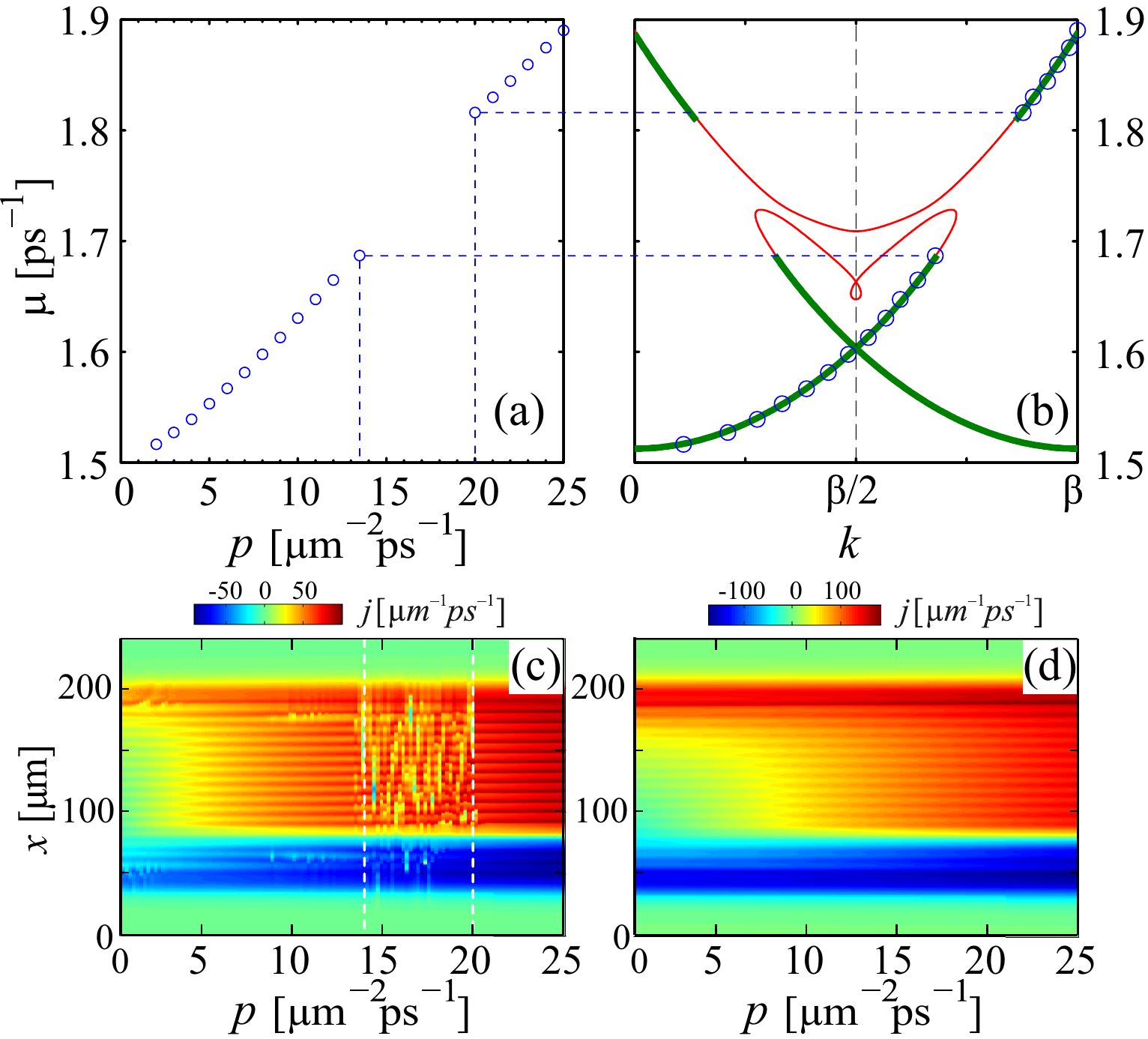}
\caption{(Color online)
The blue circles on panels (a) and (b) show the dependency of the frequency $\mu$ on the peak value of the pump $p$ for the current states forming in the finite systems with the properly designed distribution of the pump. (b) The frequency-momentum diagram plotted for the infinite system. Panels (c) and (d) illustrate the distributions of the condensate current $j$ calculated for different peak pump intensities $p$, the calculation time is  $t=1000$~ps. The pump amplitude at the plateau is $P_0=50$~$\mu$m$^{-2}$ps$^{-1}$ for panel (c) and $P_0=80$~$\mu$m$^{-2}$ps$^{-1}$ for panel (d). The potential depth is $V_0 / \hbar=0.1$~ps$^{-1}$.
}
\label{fig_curr_band}
\end{figure}

Let us briefly describe the operation principles of the suggested scheme. Higher pump intensity creates hot excitons and thus in the area of the peak of the pump the condensation rate is the highest. In the same time the hot excitons create repelling potential for the coherent polaritons. Both this factors creates a flow of polaritons from the region of the more intensive pump. That is why in the beginning of the plateau area a polariton wave with non-zero current forms. Our numerical analysis yields that on a plateau the condensate transforms into the current state provided that the pumping rate $P_0$ is appropriate for supporting a stable current in an infinite periodic lattice (see Sec.~\ref{Ch:BandStructures}). In this case condensate eventually transforms into a laminar flux of polaritons, shown in Fig.~\ref{fig_curr_dynamics}(b).

The frequency and the momentum of this current state can be attributed to the particular point of the eigenfrequency-momentum diagram of the infinite and homogeneously pumped system.
This is true for a wide range of peak amplitudes $p$ -- see blue circles in Fig.~\ref{fig_curr_band}(a) and (b).
Figure~\ref{fig_curr_band}(c) illustrates the respective distributions of a polariton current $j$ after $1000$~ps of evolution for different peak amplitudes.

For small enough values of $p$ the current profile is smooth. However the dynamics becomes turbulent for larger values of the peak amplitudes $p\gtrsim13.5$~$\mu$m$^{-2}$ps$^{-1}$.
The region of turbulence corresponds to the wide energy gap in Fig.~\ref{fig_curr_band}(a), where the nonlinear polaritons can not spread without distortions.
It turns out that the pump peak shifts the condensate energy into the energy region where only unstable current states exist [Fig.~\ref{fig_curr_band}(b)]. A typical turbulent dynamics of the condensate flow is shown in Fig.~\ref{fig_curr_dynamics}(c). With the further increase of the peak amplitude $p$ (above $p\simeq 20$~$\mu$m$^{-2}$ps$^{-1}$) the condensate jumps to the second  band, where current states are stable, and the polariton flow becomes laminar again.

The nonlinear dispersion can be gapless, see an example in panel (a) of Fig.~\ref{fig_BandStr_BZ} showing the dispersion for $P_0=80$~$\mu$m$^{-2}$ps$^{-1}$. The modifications of the current distributions for the gapless nonlinear dispersion is illustrated in panel (d) of Fig.~\ref{fig_curr_band} for $P_0=80$~$\mu$m$^{-2}$ps$^{-1}$.
In this case the flow is laminar for all values of $p$ because all the states of  nonlinear dispersion are stable.

\begin{figure}
\includegraphics[width=1\linewidth]{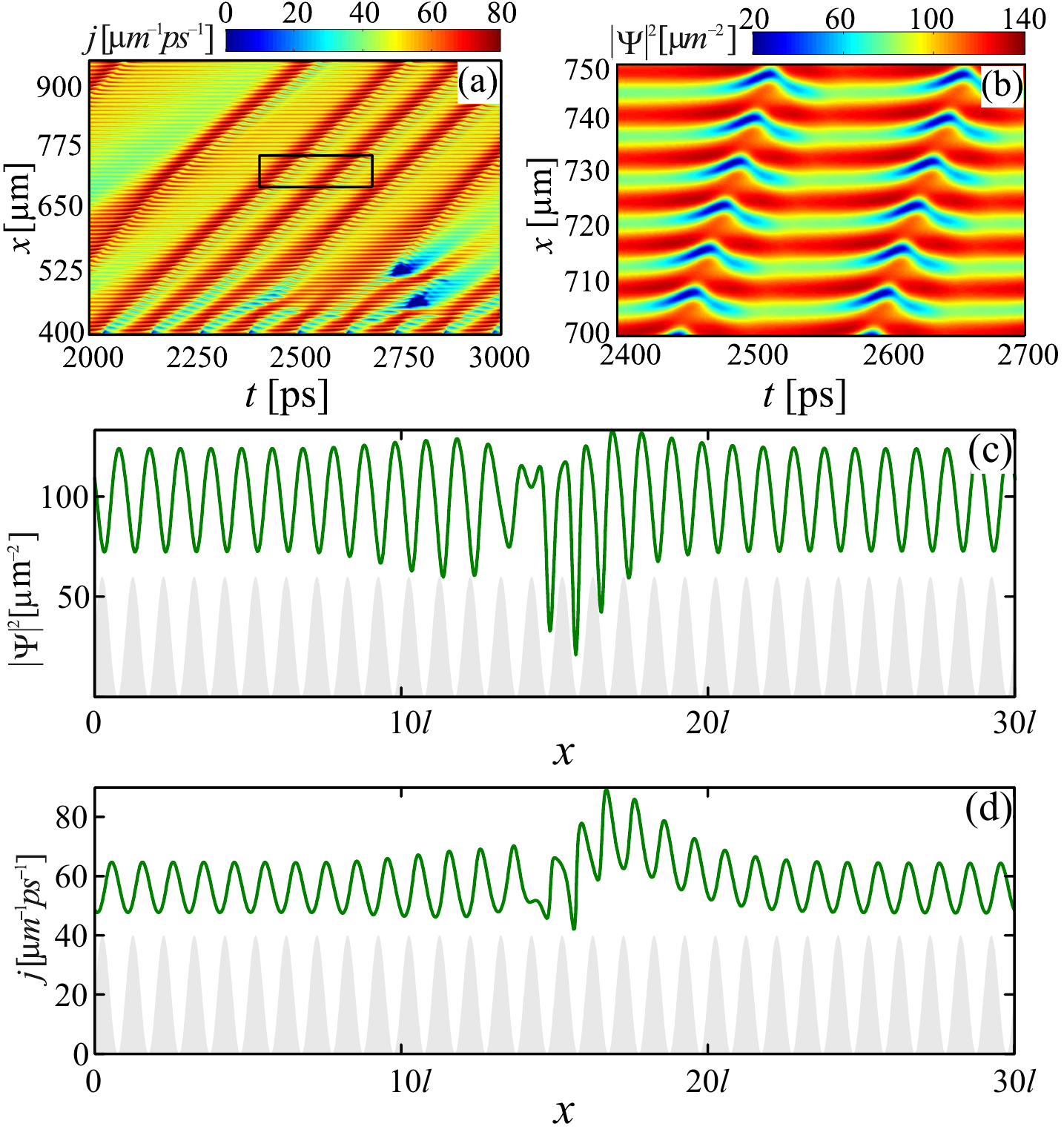}
\caption{(Color online) (a) Distribution of the polariton current $j$ illustrating the pulse propagation on the top of the current state. All parameters are the same as in Fig.~\ref{fig_curr_dynamics}(c) except for a pump plateau length (only the plateau region is shown). (b) Distribution of the condensate density $|\Psi|^2$ in the region framed in panel (a). (c) The condensate density profile $|\Psi(x)|^2$ of a solitary wave packet which remains unperturbed spreading through the long lattice. (d) the same as in panel (c) but for polariton current distribution $j(x)$.
}
\label{fig_curr_sol}
\end{figure}

In the turbulent regime the polariton condensate can spontaneously emit wave envelopes, see Fig.~\ref{fig_curr_dynamics}(c) and Fig.~\ref{fig_curr_sol} illustrating this process.
{Note that the mechanism of} emission of discussed wave packets resembles generation of dark soliton train~\cite{Pinsker2014}.
{To demonstrate that the observed pulses are dissipative solitons we performed the following numerical experiment. We cut off the parts of the solution containing the pulses from the numerical simulation of our finite system. Then we inserted these pieces of the solutions into stationary NC solution formed in the periodic system with the length much longer compared to the pulses. The parameters of the system (gain, losses, potential profile) are the same as on the plateau region of our finite system. We took these ``hybridized'' state as the initial condition and performed very long numerical simulations with periodic boundary conditions. What was observed is that these pulses propagated without any noticeable changes in their shapes.   Corresponding profiles of the condensate density distribution $|\Psi(x)|^2$ and polariton current $j(x)$  are shown in Figs.~\ref{fig_curr_sol}(c) and \ref{fig_curr_sol}(d), respectively. This gives a reason to assume that the discovered pulses are dissipative solitons.}

{Note that the proposed mechanism of formation of the polariton nonlinear current states paves the way for studying the manipulation of polariton flow and possible practical implementation of predicted phenomena in polariton circuits, cf. \cite{Liew2008, Nguyen2013,Cancellieri2015}.}

\section{Conclusions}

We performed a comprehensive theoretical analysis of nonlinear Bloch states and respective frequency-momentum dispersion for a non-equilibrium condensate of exciton-polaritons loaded into a 1D periodic lattice.
The formation of the energy loops resembling the ``swallow-tails'' of an atomic BEC was shown. {However in contrast to conservative systems the spatial structure and the frequency of Bloch solutions formed in driven-dissipative polariton system are strongly affected by the spatially distributed gain and reservoir induced effective potential. Both of these phenomena are responsible for a more complicated character of the nonlinear band-structure of exciton-polariton condensate as compared to atomic BEC.}

We demonstrated that the system allows for nonlinear current states of the polariton condensate, it is a special feature of the discussed nonlinear Bloch waves that their current does not vanish when their quasi momentum approaches the boundaries of the Brillouin zone. These current states bifurcate from the conventional nonlinear Bloch wave by the spontaneous symmetry breaking happening when the nonlinear interaction between polaritons becomes strong enough.

An analytical approach allowing to calculate the bifurcation diagrams is developed. The analytical description is approximate but the comparison between the analytically and numerically calculated diagrams shows that the accuracy of the analytical method is quite good.     It was proved for the exciton-polariton condensate that these current states occur due to effective screening of the external potential by the nonlinearity. The linear stability analysis shows that these states can be stable in a large domain of the parameters. It is shown that the states are attractors with large basins of attraction. This fact can significantly facilitate the experimental observation of the discussed states. For this we propose a special scheme allowing to observe predicted current states in real experiments.

\begin{acknowledgments}
I.Yu.Ch. acknowledges support from RFBR (grant 16-32-60102) and by the Russian Ministry of Education and Science, state task No 16.440.2014K.
A.V.Y. acknowledges support from the Government of the Russian Federation (Grant 074-U01) through the ITMO University early career fellowship.
O.A.E. acknowledges a financial support by the Deutsche Forschungsgemeinschaft (DFG project EG344/2-1).
A.P.A. and I.Yu.Ch. acknowledge financial support from RFBR, grants No. 
15-59-30406, 14-02-97503.
The financial support from the EU project (FP7, PIRSES-GA-2013-612600) LIMACONA is also acknowledged.
Fruitful discussions with Profs. Alexey Kavokin, Feo Kusmartsev and Ivan Shelykh are acknowledged.
\end{acknowledgments}

\end{document}